\documentclass{osa-article}
\usepackage{amsmath}
\usepackage{esint}
\usepackage{bm}
\usepackage{nicefrac}
\usepackage{subcaption}

\journal{oe}

\articletype{Research Article}

\newcommand\myatop[2]{\genfrac{}{}{0pt}{}{#1}{#2}} 

\begin{document}

\title{Performance Analysis of Free-space Quantum
Key Distribution Using Multiple Spatial Modes}

\author{Wenhua He,\authormark{1,*}Saikat Guha,\authormark{1,2} Jeffrey H. Shapiro,\authormark{3} and Boulat A. Bash\authormark{2,1} }

\address{\authormark{1}College of Optical Sciences, University of Arizona, Tucson AZ 85721\\
\authormark{2}Department of Electrical Engineering, University of Arizona, Tucson AZ 85721\\
\authormark{3}Research Laboratory of Electronics, Massachusetts Institute of Technology, Cambridge MA 02139}

\email{\authormark{*}whe1@optics.arizona.edu} 



\begin{abstract}
In the diffraction-limited near-field propagation regime, free-space optical quantum key distribution (QKD) systems can employ multiple spatial modes to improve their key rate. Here,  we analyze QKD using the non-orthogonal flat-top focused beams. Although they suffer from a rate penalty, their ease of implementation make them an attractive alternative to the well-studied orthonormal Laguerre-Gauss (LG) modes.  Indeed, in the presence of turbulence, the non-orthogonal modes may achieve higher QKD rate than the LG modes.
\end{abstract}

\section{Introduction}
Quantum key distribution (QKD) secures fiber and free-space optical (FSO) channels against the most powerful adversary allowed by physics \cite{scarani09rmpQKD,djordjevic19QKD,pirandola19QKDreview}.
However, its low key-rates compared to standard cryptography pose a significant challenge to its widespread adoption. The main reason for the poor performance is that the QKD {\em capacity} of a single-mode lossy optical channel, i.e., the maximum key rate attainable using any direct-transmission QKD protocol, is proportional to the end-to-end power transmissivity of the channel $\eta$. Therefore, to increase the QKD rate one must increase the number of modes used, e.g., by multiplexing over time-bins or spatial modes.
Time-bin multiplexing increases the optical bandwidth $\nu$ in modes/s while spatial-mode multiplexing employs multiple spatial modes.  
Here we investigate the latter for QKD over FSO channels.

In order to fruitfully employ multiple spatial modes over an FSO channel, that channel must support -- i.e., have appreciable transmissivities for -- multiple spatial modes.  Because QKD systems rely on dim pulses, we can expect their utility for FSO channels to be limited to clear-weather operation at wavelengths of minimal absorption.  In that case, channels of interest are subject to a modest amount of extinction loss -- from absorption and clear-weather scattering -- and the ill effects of atmospheric turbulence.  Extinction loss is a simple attenuation factor, hence we neglect it throughout this paper, so turbulence is the only atmospheric effect we consider.  To understand when multiple-spatial-mode operation is useful, it is instructive to ignore turbulence for the moment and begin with an $L$-m-long vacuum-propagation link at wavelength $\lambda$ between an area $A_{\rm T}$ transmitter pupil and an area $A_{\rm R}$ receiver pupil.  For a single polarization, that channel's power-transfer behavior has far-field and near-field regimes that are characterized by the Fresnel number product $D_{\rm f} \equiv A_{\rm T}A_{\rm R}/(\lambda L)^2$. 
In the far-field regime ($D_{\rm f} \ll 1$) only one transmitter-pupil spatial mode couples significant power into the receiver pupil (with transmissivity $\eta_1 \approx D_{\rm f}$) \cite{Sha05}, precluding appreciable improvement in the achievable QKD rate from multiple orthogonal spatial modes.
Therefore, our interest is in the {\em near-field} propagation regime ($D_{\rm f} \gg 1$). 
In this regime, approximately $D_{\rm f}$ mutually-orthogonal spatial modes have near-perfect power transmissivity ($\eta \approx 1$) \cite{Sha05}.  
These $D_{\rm f}$ modes form a signal constellation when excited one at a time, enabling $D_{\rm f}$-ary high-dimensional (HD) QKD. 
However, HD QKD's drawback is its low information density, as it improves the QKD rate only by a factor of $\log(D_{\rm f})$ \cite[Sec.~V.F]{pirandola19QKDreview}.
In principle, spatial multiplexing could improve the QKD rate by a factor of $D_{\rm f}$ by exciting all $D_{\rm f}$ modes concurrently.
We, of course, need to address FSO channels with atmospheric turbulence, as they are the channels of interest for terrestrial QKD in metropolitan areas or maritime scenarios in which fiber connections are unavailable.  The amplitude and phase fluctuations imposed on light beams propagating through turbulence can have profound effects.  Surprisingly, however, it has been shown \cite{shapiro74normalmodeturb} that the propagation geometry we specified for vacuum propagation still has far-field and near-field power-transfer regimes characterized by the same Fresnel number product.  In particular, when $D_{\rm f} \ll 1$, single-polarization propagation through turbulence has only one spatial mode with appreciable \emph{average} transmissivity $\langle n_1\rangle\approx D_{\rm f}$.  Conversely, when $D_{\rm f} \gg 1$, single-mode propagation through turbulence will support, on average, approximately $D_{\rm f}$ high-transmissivity modes.  In both regimes, however, the spatial modes of interest and their tranmissivities are, in general, random.

The orbital angular momentum (OAM) bearing Laguerre-Gauss (LG) mode set \cite{padgett17oam} has emerged as a strong candidate for spatial multiplexing in classical and quantum communication systems.
Their use in QKD was explored primarily for HD QKD \cite{mafu13hdqkdoam, mirhosseini15twistedlightqkd,intracity.3km,polarization-OAM-wang2019characterizing}, though other systems were also considered \cite{vallone14twistedlightqkd,selfhealingbessel-li2017adaptive}. LG modes remain orthonormal while propagating through vacuum \cite{Sha05}.  Each mode is thus an independent communication channel. Recent advances significantly reduce the size, weight, and cost of devices to generate and separate LG modes \cite{berkhout10oam,LG-SLM-sorter}.  Turbulence, however, destroys the orthogonality of LG modes, introducing deleterious cross-talk between them \cite{malik12oamturb,krenn14vienna}. 
The cross-talk limits their QKD rate. 

On the other hand, the use of non-orthogonal spatial modes for QKD has been limited to spatial encoding of qubits for HD QKD \cite{position-IF-walborn2006quantum,slit-etcheverry2013quantum}, which yields at most $\log(D_{\rm f})$ gain, even though such modes could capture an appreciable fraction of the $D_{\rm f}$ multiplexing gain.
Furthermore, \cite{position-IF-walborn2006quantum,slit-etcheverry2013quantum} do not account for turbulence.
We consider flat-top focused beams (FBs), which have uniform field amplitude at the transmitter pupil.
We multiplex by concurrently exciting multiple flat-top FBs that are focused on different pixels in the receiver pupil.
Thus, each beam acts as its own channel.
These modes are not orthogonal even when propagated in vacuum, and the cross-talk between the overlapping beams limits the achievable QKD rate.
Optimization of an arbitrary segmented receiver pixel tiling is computationally formidable. 
Thus, we constrain their transmitter to a square pupil, their receiver to a 100\%-fill-factor tiling of equal-area square pixels, and the protocol to decoy-state (DS) discrete-variable (DV) laser-light BB84~\cite{lo05decoyqkd}. 
This allows computation of a QKD rate-distance envelope for our flat-top FB array. 

We compare the QKD rates of our FB mode DS DV BB84 system with those of an LG-mode DS DV BB84 system.  While the FB-mode system uses hard-aperture (0 or 1 transmissivity) transmitter and receiver pupils described earlier, for the LG-mode system we assume Gaussian soft-aperture transmitter and receiver pupils that enable it to maintain orthogonality in the absence of turbulence. 
We isolate the impact of the spatial multiplexing on the QKD rate by assuming ideal photodetectors in both the LG-mode and FB systems.
To ensure fair comparison, we choose the effective areas of the soft-aperture pupils such that the LG-mode and FB-mode systems have the same Fresnel number product.
For vacuum propagation, we obtain the exact QKD rates for FB and LG mode sets. For propagation in turbulence, the exact calculation of ergodic QKD rates is prohibitively difficult.  
So, to enable tractable calculation of average power transmissivities, we use the square-law approximation in the atmospheric mutual-coherence function, which is known to be a good approximation as seen in the Appendix. 
Also, because the BB84 QKD rate is not convex in {\em both} the transmissivities and the cross-talks, we cannot obtain a rigorous lower bound. 
Hence, we resort to comparing the QKD rates as functions of average transmissivities and average cross-talks.  
We believe that the rates computed in this manner ensure an even-handed comparison between the FB and LG mode sets.
Our results, preliminary versions of which for the vacuum-propagation case we presented in~\cite{bash16multmodesQKD}, are as follows:
\begin{itemize}
    \item While, as expected, LG modes outperform flat-top FBs in vacuum, the latter capture a significant portion of the available multiplexing gain, as argued in Section \ref{sec:vacuumQKD}.
    \item Surprisingly, flat-top FBs can outperform LG modes in turbulence.
    Our results in Section \ref{sec:turbulentQKD} suggest that flat-top FBs achieve higher QKD rate than the LG modes for all but weakly-turbulent short-range links.  Furthermore, since achieving maximum rate seems to require substantially more LG modes than FBs, the LG system suffers from the additional cost of detectors and associated electronics.
\end{itemize}
Although our FB proposal is technically complicated, a similar system using an $8\times 8$ pixel receiver has been demonstrated \cite{guha14piecomm}. The
FB setup can be optimized further, e.g., using hexagonal instead of square pupils. 
Therefore, despite the many technological advances for generation and separation of LG modes \cite{berkhout10oam,LG-SLM-sorter}, concurrently-transmitted FBs offer performance/cost characteristics that suggest that LG modes might not be worth the trouble for QKD.

Next, we review the FSO channel geometry and analyze the power transmissivity for the LG modes and flat-top FBs in vacuum and under the Kolmogorov-spectrum turbulence model.  We then evaluate and compare the QKD rate attainable with these mode sets over vacuum-propagation and turbulent channels.

\section{Propagation of Light in Free-space}
\label{sec:prerequisites}
\subsection{Free-space propagation model}
\label{sec:geometry}

Consider propagation of linearly-polarized, quasimonochromatic light with center wavelength $\lambda$ from Alice's transmitter pupil in the $z=0$ plane with a field-transmission pupil function $\mathcal{A}_{\rm T}({\bm \rho})$, ${\bm \rho} \equiv (x,y)$, over an $L$-meter line-of-sight atmospheric path to Bob's receiver pupil which has field-transmission pupil function $\mathcal{A}_{\rm R}({\bm \rho^\prime})$, ${\bm \rho^\prime} \equiv (x^\prime,y^\prime)$.
Alice's transmitted field's complex envelope $E_0({\bm \rho},t)$ is multiplied by $\mathcal{A}_{\rm T}({\bm \rho})$, undergoes free-space diffraction and turbulence over the $L$-meter path, and is truncated by $\mathcal{A}_{\rm R}({\bm \rho^\prime})$, to yield the received field $E_L({\bm \rho^\prime},t)$. 
Neglecting extinction, these input and output field envelopes are related by the extended Hugyens-Fresnel principle that, because turbulence is non-depolarizing, with $\approx$THz coherence bandwidth, and $\approx$ms coherence time can be taken to be: 
\begin{align}
E_L({\bm \rho^\prime},t)&=  \int E_0({\bm \rho},t - L/c) \, h({\bm \rho^\prime}, {\bm \rho}, t) \, \mathrm{d}^2 {\bm \rho}
\end{align}
for QKD pulse durations $\gg$ps, where $c$ is the speed of light.
The integrals in this paper are over the plane $\mathbb{R}^2$ unless otherwise indicated.
Moreover, for QKD pulse durations $\ll$ms we can drop the time argument $t$ from the impulse response $h({\bm \rho^\prime}, {\bm \rho}, t)$ and treat it statistically.  

For vacuum propagation, the extended Huygens-Fresnel principle reduces to the ordinary Huygens-Fresnel principle and yields: 
\begin{align}
\label{eq:vacpropkernel}  h_{\rm vac}({\bm \rho^\prime}, {\bm \rho})& =\mathcal{A}_{\rm T}({\bm \rho^\prime})  \frac{\exp\left[ik \left(L + |{\bm \rho^\prime}-{\bm \rho}|^2/2L\right)\right]}{i \lambda L} \mathcal{A}_{\rm R}({\bm \rho}),
\end{align}
where $k \equiv 2\pi/\lambda$  is the wave number and we have folded the field-transmission pupil functions into the spatial impulse response.
Let the transmitted field be $E_0(\bm{\rho})=\sqrt{P_{\rm T}}u_0(\bm{\rho})$, where the mode pattern satisfies $\int_{\mathbb{R}^2}|u_0(\bm{\rho})|^2\mathrm{d}^2 {\bm \rho}=1$ and $P_{\rm T}$ is the transmitted power in photons/s.
Singular value decomposition of the impulse response $h({\bm \rho^\prime}, {\bm \rho})=\sum_{q=1}^\infty\sqrt{\eta_q}\phi_q(\bm{\rho}^{\prime})\Phi_q^*(\bm{\rho})$ yields a complete orthonormal (CON) set of functions (input modes) $\{\Phi_q(\bm{\rho})\}$ and the corresponding CON set of functions (output modes) $\{\phi_q(\bm{\rho})\}$, where $1\geq\eta_1\geq\eta_2\geq\ldots\geq 0$ are the modal transmissivities.
That is, transmission of $u_0({\bm \rho^\prime})=\Phi_q(\bm{\rho})$ results in reception of $E_L({\bm \rho^\prime})=\sqrt{\eta_q}\phi_q(\bm{\rho})$, implying that the channel $h({\bm \rho^\prime}, {\bm \rho})$ can be decomposed into a countably infinite set of parallel channels.
The subset of these channels that is useful for spatial multiplexing is characterized by the Fresnel number product
\begin{align}
\label{eq:Df}
    D_{\rm f}&=\iint\langle|h({\bm \rho^\prime}, {\bm \rho})|^2\rangle\mathrm{d}^2{\bm \rho^\prime}\mathrm{d}^2{\bm \rho}=\sum_{q=1}^\infty \langle \eta_q\rangle =\frac{A_{\rm T}A_{\rm R}}{(\lambda L)^2},
\end{align}
where  $A_{\rm T}=\int\left|\mathcal{A}_{\rm T}(\bm{\rho})\right|^2\mathrm{d}^2 \bm{\rho}$ and $A_{\rm R}=\int\left|\mathcal{A}_{\rm R}(\bm{\rho^\prime})\right|^2\mathrm{d}^2 \bm{\rho^\prime}$ are the transmitter and receiver pupil areas.
For simplicity of exposition, we assume these to be equal, $A_{\rm T}=A_{\rm R}=A$.
In the \emph{near-field} regime where $D_{\rm f}\gg 1$, $\approx D_{\rm f}$ modes couple significant power between transmitter and receiver, allowing spatial multiplexing.
In the \emph{far-field} regime where $D_{\rm f}\ll 1$, only one mode has significant transmissivity $\eta\approx D_{\rm f}$.

We model turbulent propagation using Kolmogorov-spectrum turbulence with zero inner scale, infinite outer scale, and uniform strength $C_n^2$ from $z=0$ to $z=L$.
The extended Huygens-Fresnel principle's impulse response can be conveniently written as:
\begin{align}
\label{eq:turbpropkernel} h({\bm \rho^\prime}, {\bm \rho})& = h_{\rm vac}({\bm \rho^\prime}, {\bm \rho})e^{\psi({\bm \rho^\prime}, {\bm \rho})},
\end{align}
where $\psi({\bm \rho^\prime}, {\bm \rho})=\chi({\bm \rho^\prime}, {\bm \rho})+i\phi({\bm \rho^\prime}, {\bm \rho})$, with $\chi({\bm \rho^\prime}, {\bm \rho})$ and $\phi({\bm \rho^\prime}, {\bm \rho})$ being the random log-amplitude and phase fluctuations produced by the turbulence.
In the analysis that follows we employ the atmospheric impulse response mutual coherence function \cite{shapiro74normalmodeturb,shapiro78turbNoAuthorNoTitle,lutomirski71inhom,ishimaru78waveprop}:
\begin{align}
    \label{eq:mutualcoherence}\langle h({\bm \rho^\prime_1}, {\bm \rho_1})h^\ast({\bm \rho^\prime_2}, {\bm \rho_2})\rangle&=h_{\rm vac}({\bm \rho^\prime_1}, {\bm \rho_1})h_{\rm vac}^\ast({\bm \rho^\prime_2}, {\bm \rho_2}) e^{-\frac{1}{2}D\left(\bm{\rho}_{1}^{\prime}-\bm{\rho}_{2}^{\prime}, \bm{\rho}_{1}-\bm{\rho}_{2}\right)},
\end{align}
where $D(\Delta\bm{\rho}^{\prime},\Delta\bm{\rho})$ is the two-source, spherical-wave, wave structure function
\begin{align}
    \label{eq:D}D(\Delta\bm{\rho}^{\prime},\Delta\bm{\rho})&=2.91k^2C_n^2\int_0^L\left|\Delta\bm{\rho}^{\prime}z/L+\Delta\bm{\rho}(1-z/L)\right|^{5/3}\mathrm{d}z.
\end{align}
The integral in \eqref{eq:D} significantly complicates the analysis.
Thus, instead of the $5/3$-law structure function, we employ its square-law approximation \cite{chandrasekaran14pieturb1, 2014-crosstalk-simulator}
 \begin{align}
\label{eq:Dsq}D_{\mathrm{sq}}(\Delta \bm{\rho}^{\prime}, \Delta \bm{\rho})&=\frac{\left|\Delta \bm{\rho}^{\prime}\right|^{2}+\Delta \bm{\rho}^{\prime} \cdot \Delta \bm{\rho}+|\Delta \bm{\rho}|^{2}}{\rho_{0}^{2}},
 \end{align}
where $\rho_{0}=\left(1.09 k^{2} C_{n}^{2} L\right)^{-3/5}$ is the spherical-wave coherence length.
Appendix \ref{wave-structure-function} shows that \eqref{eq:Dsq} is a fairly accurate approximation of \eqref{eq:D} for power-transfer calculations. 

\subsection{Orthogonal mode sets for soft Gaussian pupils}
\label{sec:gaussian_pupils}

\begin{figure}[h]
     \centering
     \begin{subfigure}[b]{0.49\textwidth}
         \centering
         \includegraphics[width=\textwidth]{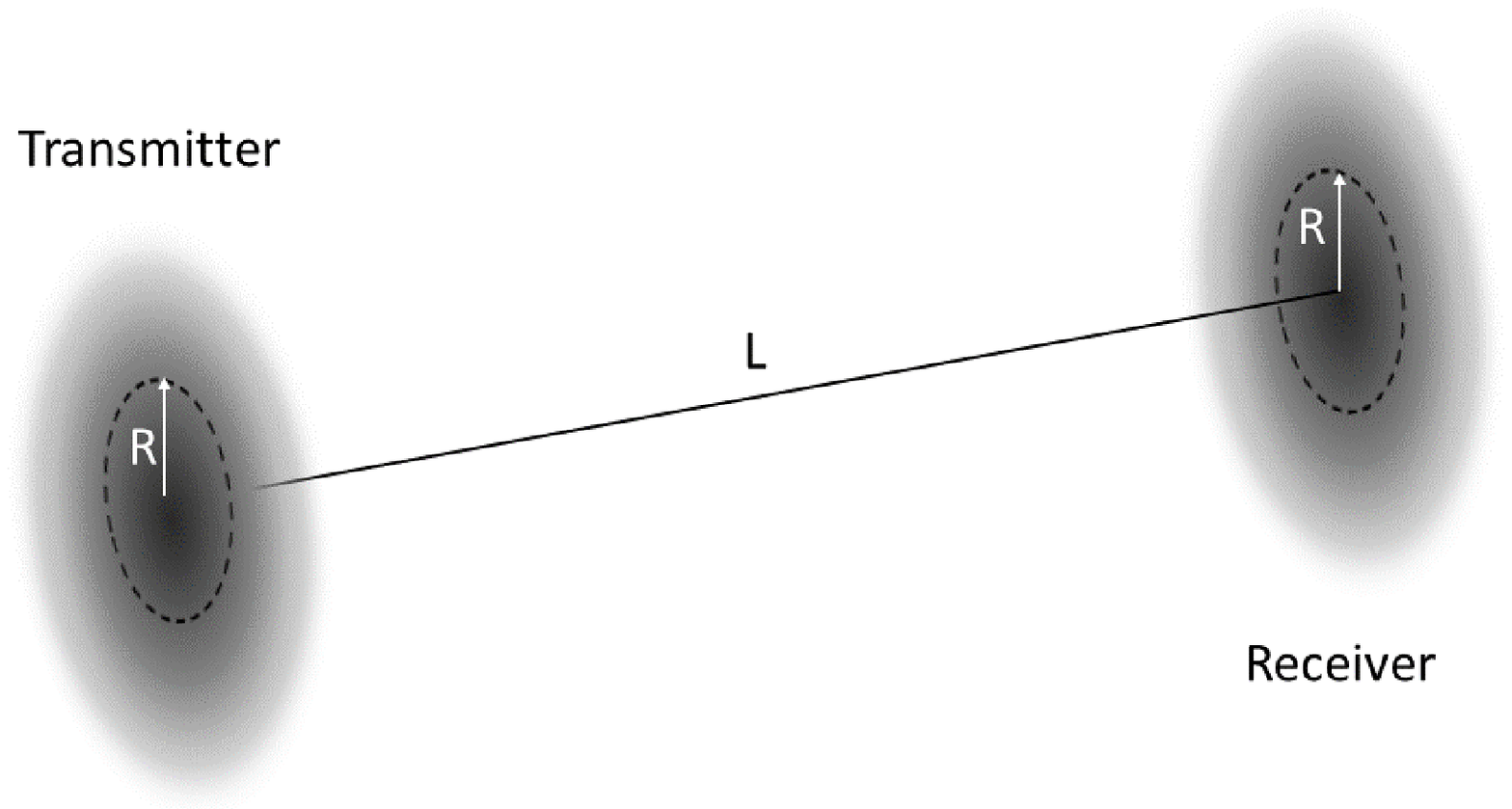}
         \caption{Soft Gaussian pupil geometry.}\label{fig:soft}
     \end{subfigure}
     \hfill
     \begin{subfigure}[b]{0.49\textwidth}
         \centering
         \includegraphics[width=\textwidth]{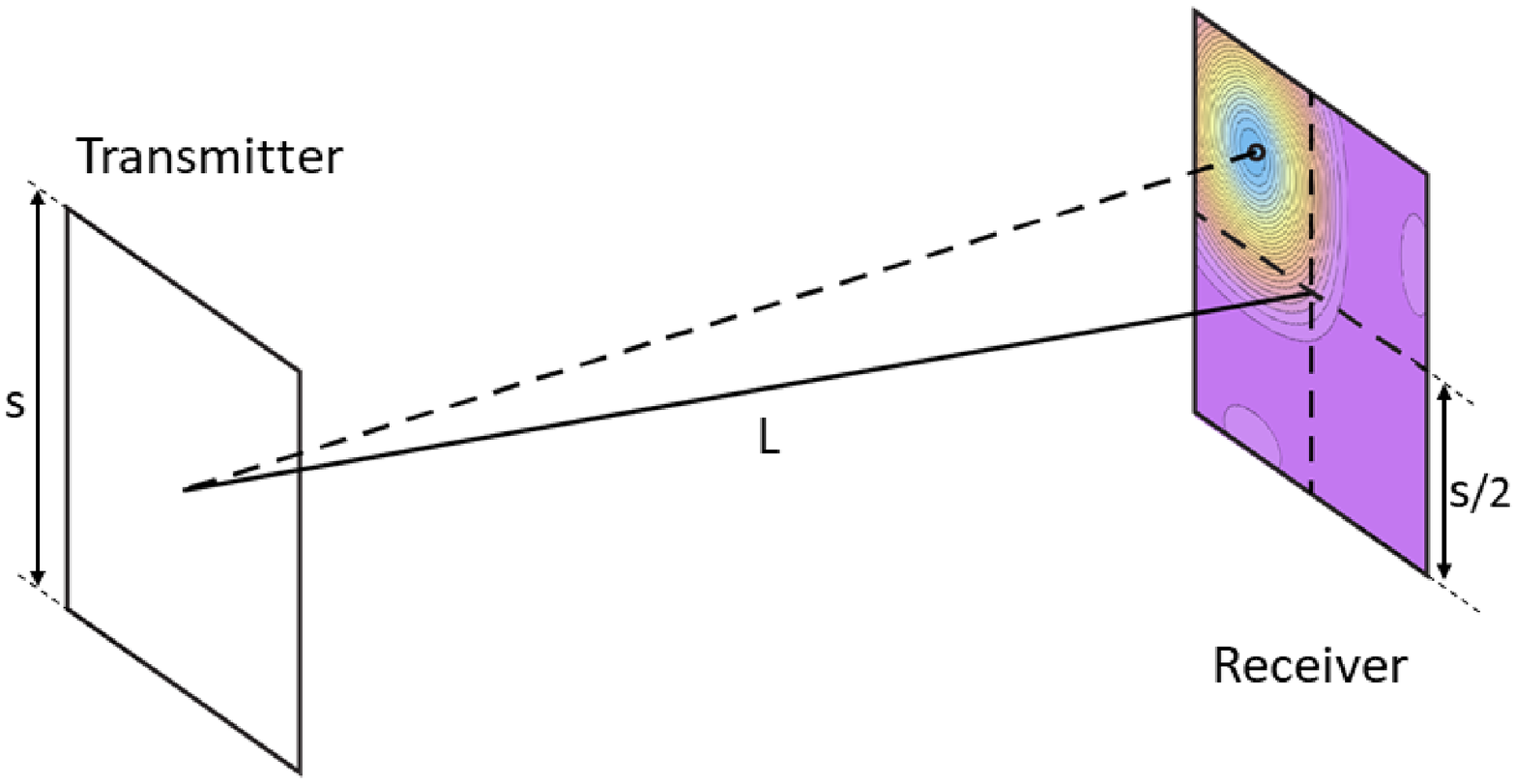}
         \caption{Unapodized square pupil geometry.}\label{fig:hard}	
     \end{subfigure}
        \caption{Line-of-sight system geometries. Soft Gaussian pupils in (\subref{fig:soft}) support orthogonal propagation of LG modes in vacuum.  An example of unapodized square pupil system in (\subref{fig:hard}) uses four FBs focused on a $2 \times 2$-pixel receiver. Irradiance pattern from one of the FBs illustrates the cross-talk.  To compare LG and FB mode sets fairly, we make their respective Fresnel number products equal.}
     \label{fig:line-of-sight-system}
\end{figure}

Consider the vacuum-propagation impulse response $h_{\rm vac}({\bm \rho^\prime}, {\bm \rho})$ in \eqref{eq:vacpropkernel} and the soft Gaussian pupil channel geometry depicted in Figure \ref{fig:line-of-sight-system}(\subref{fig:soft}).  The pupils' pupil function is:
\begin{align}
    \label{eq:gaussian_pupils}\mathcal{A}(\bm{\rho})=\exp\left[-\frac{\left|\bm{\rho}\right|^2}{R^2}\right],
\end{align}
where $R$ is pupil effective radius.
The area of the Gaussian pupil is $A=\frac{\pi R^2}{2}$.
LG modes are labeled by the radial and azimuthal indices $p=0,1,2,\ldots $ and  $l=0,\pm 1,\pm 2, \ldots$.
The input LG mode indexed by  $\mathbf{q}\equiv(p,l)$ is expressed using the polar coordinates ${\bm \rho}\equiv(r,\theta)$ as \cite[Section 3.A]{Sha05}:
\begin{align}
\label{eq:inputLG} \Phi_{\bm{q}}^{\rm (LG)}(r,\theta)&=\sqrt{\frac{p!}{\pi(|l|+p)!}}\frac{1}{A_{\rm T}}\left[\frac{r}{A_{\rm T}}\right]^{|l|}\mathcal{L}_p^{|l|}\left(\frac{r^2}{A_{\rm T}^2}\right)\exp\left(-\left[\frac{1}{2A_{\rm T}^2}+\frac{ik}{2L}\right]r^2+il\theta\right),
\end{align}
where $\mathcal{L}_p^{|l|}(\cdot)$ denotes the generalized Laguerre polynomial 
  indexed  by $p$ and $|l|$, and $A_{\rm T}$ is the transmitter pupil area (set to $A_{\rm T}=A$ in this paper).
The corresponding output LG mode is:
\begin{align}
\label{eq:outputLG} \phi_{\bm{q}}^{\rm (LG)}(\bm{\rho}^\prime)&=\sqrt{\frac{p!}{\pi(|l|+p)!}}\frac{1}{i^{2p+|l|+1}A_{\rm R}}\left[\frac{r^\prime}{A_{\rm R}}\right]^{|l|}\mathcal{L}_p^{|l|}\left(\frac{(r^\prime)^2}{A_{\rm R}^2}\right)\exp\left(-\left[\frac{1}{2A_{\rm R}^2}-\frac{ik}{2L}\right](r^\prime)^2+il\theta^\prime\right),
\end{align}
where $\bm{\rho}^\prime\equiv(r^\prime,\theta^\prime)$, and $A_{\rm R}$ is the receiver pupil area (set to $A_{\rm R}=A$ in this paper).
The modal transmissivities are:
\begin{align}
\eta_{\bm q}^{\rm (vac)}& =\left(\frac{1+2D_\mathrm{f}-\sqrt{1+4D_\mathrm{f}}}{2D_\mathrm{f}}\right)^{q},
\label{eq:gen_modes}
\end{align}
where $D_{\rm f}=\left(\frac{kR^2}{4L}\right)^2$ is the Fresnel number product defined in \eqref{eq:Df} and $q=2p+|l|+1$ is the mode order.
Note that in the far-field regime ($D_{\rm f}\ll 1$), the focused Gaussian beam (${\bm q}={\bm 0}$) has transmissivity $\eta_{\bm 0}^{\rm (vac)}\approx D_{\rm f}$, while other modes' transmissivities are insignificant. 

Let an LG mode $\Phi_{\mathbf{q}}^{\rm (LG)}(\bm{\rho})$,  $\mathbf{q}\equiv(p,l)$, be transmitted over a vacuum-propagation channel with the impulse response in \eqref{eq:vacpropkernel} and soft pupils in \eqref{eq:gaussian_pupils} at both transmitter and receiver.
Denote by $E^{\rm (vac)}_{\mathbf{q}}(\bm{\rho}^\prime)$ the field that emerges from the receiver's pupil.
Suppose we use an ideal mode converter to extract spatial mode $\phi_{\mathbf{q}^\prime}^{\rm (LG)}(\bm{\rho}^\prime)$, $\mathbf{q}^\prime\equiv(p^\prime,l^\prime)$, from the $E^{\rm (vac)}_{\mathbf{q}}(\bm{\rho}^\prime)$.
We call power-in-fiber the power collected when such mode converter is used to direct the power from $\phi_{\mathbf{q}^\prime}^{\rm (LG)}(\bm{\rho}^\prime)$ into the propagating mode of a single-mode fiber.
Since LG modes remain orthogonal in vacuum propagation, the power-in-fiber is:
\begin{align}
    P_{\mathbf{q}\mathbf{q}^\prime}^{\rm (vac)}&\equiv\left|\int_{\mathbb{R}^2}E^{\rm (vac)}_{\mathbf{q}}(\bm{\rho}^\prime)\left[\phi_{\mathbf{q}^\prime}^{\rm (LG)}(\bm{\rho}^\prime)\right]^*\mathrm{d}^2\bm{\rho}^\prime\right|^2=\left\{\begin{array}{ll}P_{\rm T}\eta_{\bm q}^{\rm (vac)},&\bm{q}=\bm{q}^\prime\\0,&\text{otherwise.}\end{array}\right.
\end{align}
The total power captured by the receiver pupil, or the power-in-bucket, is then:
\begin{align}
\label{eq:etaG_PIB}    P_{\mathbf{q}\to\mathrm{R}}^{\rm (vac)}&\equiv \int_{\mathbb{R}^2}\left|E^{\rm (vac)}_{\mathbf{q}}(\bm{\rho}^\prime)\right|^2\mathrm{d}^2\bm{\rho}^\prime=\sum_{\mathbf{q}^\prime}P_{\mathbf{q}\mathbf{q}^\prime}^{\rm (vac)}=P_{\rm T}\eta_{\bm q}^{\rm (vac)}.
\end{align}

Now consider transmission of $\Phi_{\mathbf{q}}^{\rm (LG)}(\bm{\rho})$ over a turbulent channel with the impulse response in \eqref{eq:turbpropkernel}, and denote by $E_{\mathbf{q}}(\bm{\rho}^\prime)$ the field that emerges from the receiver's pupil.
Turbulence destroys the orthogonality of LG modes.
The power-in-fiber from coupling $\phi_{\mathbf{q}^\prime}^{\rm (LG)}(\bm{\rho}^\prime)$ mode of $E_{\mathbf{q}}(\bm{\rho}^\prime)$, averaged over the random fluctuations from turbulence, is:
\begin{align}
    \langle P_{\mathbf{q}\mathbf{q}^\prime}^{\rm (LG)}\rangle&\equiv\left\langle\left|\int_{\mathbb{R}^2}E_{\mathbf{q}}(\bm{\rho}^\prime)\left[\phi_{\mathbf{q}^\prime}^{\rm (LG)}(\bm{\rho}^\prime)\right]^*\mathrm{d}^2\bm{\rho}^\prime\right|^2\right\rangle=P_{\rm T}\langle\eta_{\mathbf{q}\mathbf{q}^\prime}\rangle.
\end{align}
Note that $\langle\eta_{\mathbf{q}\mathbf{q}}\rangle$ is the average fraction of power in $\Phi_{\mathbf{q}}^{\rm (LG)}(\bm{\rho})$ that couples to $\phi_{\mathbf{q}}^{\rm (LG)}(\bm{\rho})$ at the receiver, while $\langle\eta_{\mathbf{q}\mathbf{q}^\prime}\rangle$, $\mathbf{q}\neq\mathbf{q}^\prime$ is the average undesired cross-talk introduced by the turbulence.
The average power-in-bucket is:
\begin{align}
    \langle P_{\mathbf{q}\to\mathrm{R}}^{\rm (LG)}\rangle&\equiv \left\langle\int_{\mathbb{R}^2}\left|E_{\mathbf{q}}(\bm{\rho}^\prime)\right|^2\mathrm{d}^2\bm{\rho}^\prime\right\rangle=\sum_{\mathbf{q}^\prime}\langle P_{\mathbf{q}\mathbf{q}^\prime}^{\rm (LG)}\rangle\geq \langle P_{\mathbf{q}\mathbf{q}}^{\rm (LG)}\rangle
\end{align}
since $\{\phi_{\mathbf{q}}^{\rm (LG)}(\bm{\rho})\}$ is a CON set.
In the far-field we only transmit the focused Gaussian beam and do not employ a mode sorter.
We attain an average power-in-bucket $\langle P_{{\bm 0}\to\mathrm{R}}^{\rm (LG)}\rangle=P_{\rm T}\langle\eta_{{\bm 0}\to\mathrm{R}}\rangle$, where:
\begin{align}
    \langle\eta_{{\bm 0}\to\mathrm{R}}\rangle&=\eta^{\rm (vac)}_{\bm 0}\frac{1+4 D_{\rm f}+\sqrt{1+4 D_{\rm f}}}{1+4 D_{\rm f}+\sqrt{1+4 D_{\rm f}}+\left(R/\rho_{0}\right)^{2}}.
\end{align}
In the far-field regime ($D_{\rm f}\ll 1$) when turbulence dominates the diffraction $\rho_0 \ll R$,  $\langle\eta_{{\bm 0}\to\mathrm{R}}\rangle\approx\frac{2D_{\rm f}\rho_0^2}{R^2}$.
Thus, turbulence changes how the far-field transmissivity scales with path length $L$: from $\eta_0^{\rm (vac)}\propto L^{-2}$ in vacuum to $\langle\eta_{{\bm 0}\to\mathrm{R}}\rangle\propto L^{-16/5}$.

The exact expression for average $\langle\eta_{\mathbf{q}\mathbf{q}^\prime}\rangle$ under our zero inner scale, infinite outer scale, and uniform strength Kolmogorov-spectrum turbulence model is:
\begin{align}
\label{eq:turb_eta}\langle\eta_{\mathbf{q}\mathbf{q}^\prime}\rangle &=\iiiint \phi^{*}_{\mathbf{q}^\prime}(\bm{\rho}_{1}^{\prime}) \phi_{\mathbf{q}^\prime}(\bm{\rho}_{2}^{\prime}) \langle h({\bm \rho^\prime_1}, {\bm \rho_1})h^\ast({\bm \rho^\prime_2}, {\bm \rho_2})\rangle \Phi_{\mathbf{q}}(\bm{\rho}_{1}) \Phi_{\mathbf{q}}^{*}(\bm{\rho}_{2})\mathrm{d}^2\bm{\rho}_{1}^{\prime}\mathrm{d}^2 \bm{\rho}_{2}^{\prime}\mathrm{d}^2\bm{\rho}_{1}\mathrm{d}^2\bm{\rho}_{2},
\end{align}
where the mutual coherence function $\langle h^\ast({\bm \rho^\prime_1}, {\bm \rho_1})h({\bm \rho^\prime_2}, {\bm \rho_2})\rangle$ is given in \eqref{eq:mutualcoherence}.
Direct numerical evaluation of \eqref{eq:turb_eta} using \eqref{eq:D} is a daunting task.
The latter contains an integral that requires numerical evaluation, and it is embedded in an 8-dimensional integral that also must be evaluated numerically. 
However, the LG modes are related to the rectangularly-symmetric Hermite-Gauss (HG) modes via a unitary transformation \cite[Eq.~(8), (9)]{beijersbergen93modeconverters}.
The square-law approximation in \eqref{eq:Dsq} allows exploiting HG modes' symmetry to reduce the evaluation of \eqref{eq:turb_eta} to a product of two 4-dimensional integrals.
We then employ the unitary in \cite[Eq.~(8), (9)]{beijersbergen93modeconverters} to calculate the square-law approximation of $\langle\eta_{\mathbf{q}\mathbf{q}^\prime}\rangle$ for the LG modes.

\subsection{Flat top focused beam array}
\label{sec:square_pupils}
Practical systems employ hard transmitter and receiver pupils. 
Consider the channel geometry in Figure \ref{fig:line-of-sight-system}(\subref{fig:hard}) that employs unapodized (hard) $s\times s$ m square pupils:
\begin{align}
  \mathcal{A}(x,y)&=\begin{cases}
    1,& \text{if } |x|\leq \frac{s}{2} \text{~and~} |y| \leq \frac{s}{2},\\
    0,              & \text{otherwise}.
\end{cases}
\end{align}
For vacuum propagation using hard circular or rectangular pupils, prolate-spheroidal functions form the CON mode sets \cite{slepian64prolate,slepian65apodization}, however, their complexity renders them impractical.
Instead, we transmit uniform-fill (flat-top) beams $\Phi_{(n,m)}^{\rm (FB)}(\bm{\rho})$, $n=1,\ldots,N$, $m=1,\ldots,N$ focused on one of the equal-area square pixels in the $N\times N$ or $2\times 1$ 100\% fill-factor receiver array:
\begin{align}
    \Phi_{(n,m)}^{\rm (FB)}(\bm{\rho})&= \frac{1}{s}\exp\left\{-\frac{ik}{2L} \left[\left(x-x^\prime_{n}\right)^2+\left(y-y^\prime_{m}\right)^2\right]\right\},
\end{align}
where $\left(x^\prime_{n},y^\prime_{m}\right)$ is the center coordinate of pixel $(n,m)$.
The power-in-bucket from focused beam $\Phi_{(n,m)}^{\rm (FB)}(\bm{\rho})$ to pixel $(n^\prime,m^\prime)$ in vacuum is $P^{\rm (vac, FB)}_{(n,m)\to(n^\prime,m^\prime)}=P_{\rm T}\eta^{\rm (vac, FB)}_{(n,m)\to(n^\prime,m^\prime)}$, where 
\begin{align}
    \label{eq:etaFBvac}\eta^{\rm (vac, FB)}_{(n,m)\to(n^\prime,m^\prime)}&= I^{\rm (vac)}(x^\prime_{n}-x^\prime_{n^
    \prime})I^{\rm (vac)}(y^\prime_{m}-y^\prime_{m^
    \prime})
    \end{align}
and
    \begin{align}
 I^{\rm (vac)}(a)&=\frac{\sqrt{D_{\rm f}}}{N}\int_{a N/s-1/2}^{a N/s+1/2}\operatorname{sinc}^2\left(\frac{\pi  s^2 \xi}{N \lambda L }\right)\mathrm{d} \xi.
\end{align} 
Here, the Fresnel number product is $D_{\rm f}=\left(\frac{s^2}{\lambda L}\right)^2$.
The example of a $2\times2$-pixel focused beam array in Figure \ref{fig:line-of-sight-system}(\subref{fig:hard}) illustrates the cross-talk between the focused beams.
We match the area of the hard pupils to the area of soft Gaussian pupils in Section \ref{sec:gaussian_pupils} by setting $s=\frac{\sqrt{\pi}R}{\sqrt{2}}$.
This matches the far-field power transmissivity in vacuum, which is $\approx D_{\rm f}$ for both designs.

\begin{figure}[htbp]
\centering
\includegraphics[width=0.6\linewidth]{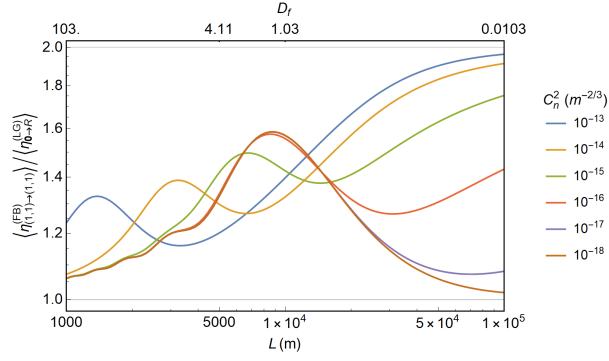}
\caption{Comparison of the average power transmissivities $\langle\eta^{\rm (FB)}_{(1,1)\to(1,1)}\rangle$ and $\langle\eta_{{\bm 0}\to\mathrm{R}}^{\rm (LG)}\rangle$ for flat-top and Gaussian focused beams at various turbulence strengths. The pupil areas are  $A_{\rm T}=A_{\rm R}=157 \text{~cm}^2$ and wavelength is $\lambda=1.55 ~\mu$m.}
\label{fig:powertransfertwice}
\end{figure}

In turbulence the average power-in-bucket from focused beam $\Phi_{(n,m)}^{\rm (FB)}(\bm{\rho})$ to pixel $(n^\prime,m^\prime)$ is $\langle P^{\rm (FB)}_{(n,m)\to(n^\prime,m^\prime)}\rangle=P_{\rm T}\langle\eta^{\rm (FB)}_{(n,m)\to(n^\prime,m^\prime)}\rangle$, where 
\begin{align}
    \label{eq:etaturbFB}\langle\eta^{\rm (FB)}_{(n,m)\to(n^\prime,m^\prime)}\rangle&=I(n,n^\prime) I(m,m^\prime)
\end{align}
and
\begin{align}
    I(n,n^\prime)&=\frac{2 \sqrt{D_{\rm f}}}{N} \int_{0}^{1}  (1-\xi) \operatorname{sinc}\left(\frac{\pi s^2 \xi}{N\lambda L}\right) \exp \left\{-\frac{\xi^{2} s^{2}}{2 \rho_{0}^{2}}\right\}\cos\left(\frac{2\pi s^2 \xi}{N\lambda L}\left(n^\prime-n\right)\right)\mathrm{d}\xi.
\end{align}
For the single-pixel receiver ($N=1$), the average transmissivity reduces to
\begin{align}
       \langle\eta^{\rm (FB)}_{(1,1)\to(1,1)}\rangle&=4 D_{\rm f}\left[ \int_{0}^{1}  \operatorname{sinc}\left(\pi \sqrt{D_{\rm f}} \xi\right) \exp \left(-\frac{\xi^{2} s^{2}}{2 \rho_{0}^{2}}\right)\mathrm{d} \xi (1-\xi)\right]^{2}.
        \label{flat top FBfar}
\end{align}
In the far-field regime ($D_{\rm f}\ll 1$) and when turbulence dominates the diffraction $\rho_0 \ll R$,  $\langle\eta_{(1,1)\to(1,1)}^{\rm (FB)}\rangle\approx\frac{2\pi D_{\rm f}\rho_0^2}{s^2}$.
Thus, the power transmissivity for the focused beam through hard square pupils in the far field has the same $L^{-16/5}$ scaling as that for the Gaussian beam through soft Gaussian pupils.
However, when the areas of the respective pupils match, 
$\lim_{L\to\infty}\nicefrac{\langle\eta^{\rm (FB)}_{(1,1)\to(1,1)}\rangle}{\langle\eta_{{\bm 0}\to\mathrm{R}}^{\rm (LG)}\rangle}=2$,
indicating that the flat-top focused beams are more resilient to turbulence than the focused Gaussian beams in the far field.
Further evidence of this resilience is provided in Figure \ref{fig:powertransfertwice}, which illustrates the superiority of the flat-top focused beams over the focused Gaussian beam in the near-field as well. 

\section{Spatially-multiplexed Quantum Key Distribution}\label{QKDratecalcmethod}
We now employ the methodology developed in Section \ref{sec:prerequisites} to analyze spatially-multiplexed QKD.
After describing our system setup, we show that the orthogonality of the LG modes allows them to outperform flat-top focused beams in vacuum, although the latter does capture a significant portion of the possible multiplexing gain.
We then show that even though turbulence increases the cross-talk between the focused beams and, thus, degrades their QKD rate, they outperform the LG modes.

\subsection{System setup}
We set the radius of soft pupils to $R=10$ cm.  Thus, the matching side length of the hard square pupil is $s=12.53$ cm, and area $A=157\text{~cm}^2$. 
We numerically evaluate the rate of decoy state (DS) discrete-variable (DV) QKD protocol \cite{lo05decoyqkd} for line-of-sight propagation of laser light at $\lambda=1.55~\mu$m center-wavelength over path lengths $L\in [1,100]$ km.
We let the probability that the pulse polarization is maintained between preparation of the polarized light pulse by Alice and its measurement by Bob (called ``visibility'' in \cite{scarani09rmpQKD}) $V=0.99$.
We assume a $\nu=10^{10}$ Hz optical bandwidth, unity detector quantum efficiency, and the availability of capacity-achieving error correction codes.
We lump the background light and dark counts together into an effective dark-count probability $p_{\rm dc}=10^{-6}$.
We treat the erroneous counts from cross-talk as additional detector dark counts.
Therefore, the QKD rate $\mathcal{R}_{\rm QKD}(\eta_{\bm q},P_{\bm q}^{\rm (T)},P_{\bm q}^{\rm (C)})$ for mode ${\bm q}$ depends on its transmissivity $\eta_{\bm q}$, the transmitted power $P_{\bm q}^{\rm (T)}$ allocated to it, and the power $P_{\bm q}^{\rm (C)}$ of the cross-talk received from other modes \cite[Sec.~IV.B.3]{scarani09rmpQKD}, \cite{lo05decoyqkd}.

\subsection{QKD over vacuum-propagation channels}
\label{sec:vacuumQKD}
\begin{figure}[t] 
    \centering
\begin{subfigure}[t]{0.49\linewidth}
		\centering
		\includegraphics[width=\linewidth]{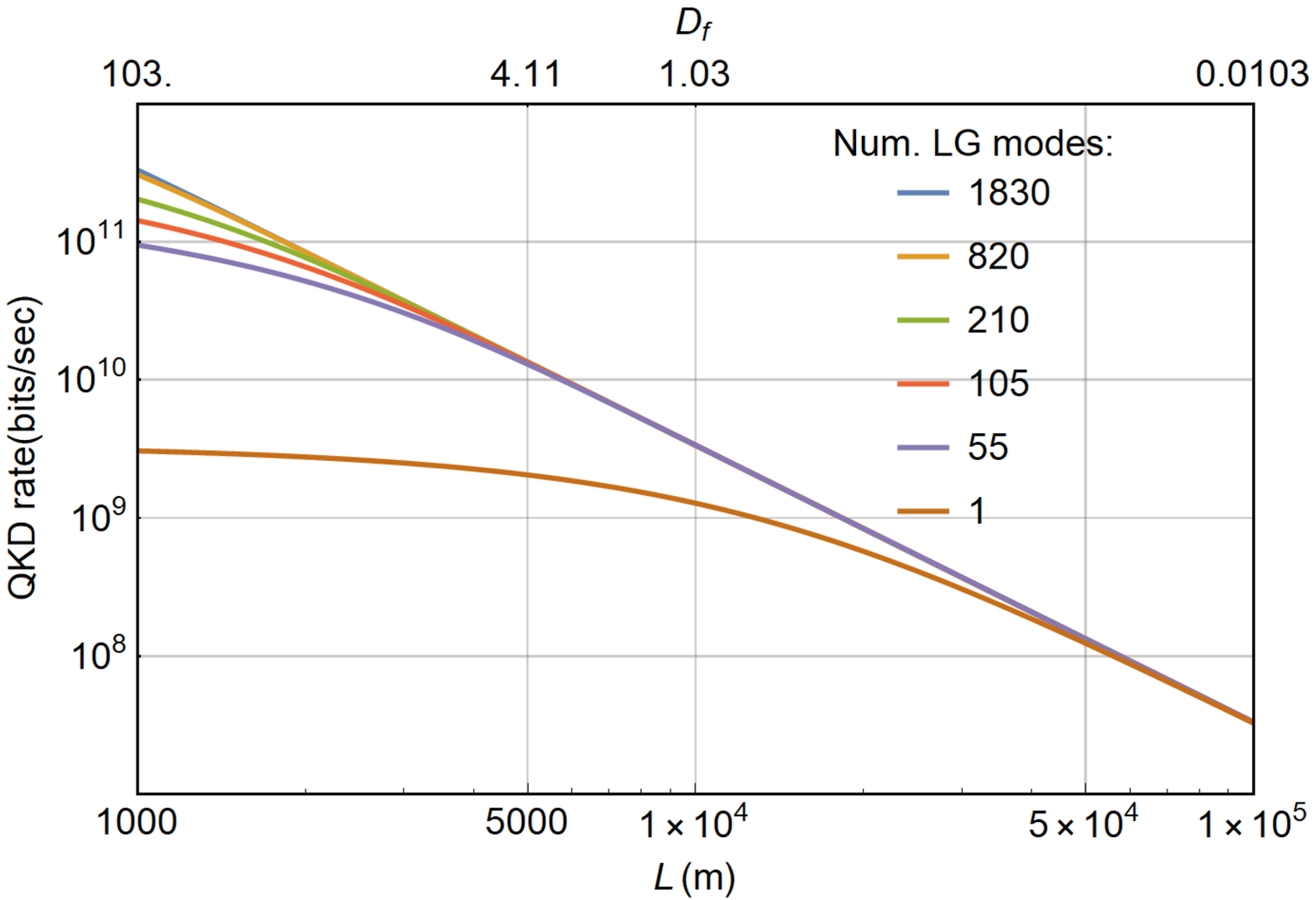}
		\caption{LG modes.}\label{fig:vacuum-LG-orthogonal}		
	\end{subfigure}
	\hfill
\begin{subfigure}[t]{0.49\linewidth}
		\centering
		\includegraphics[width=\linewidth]{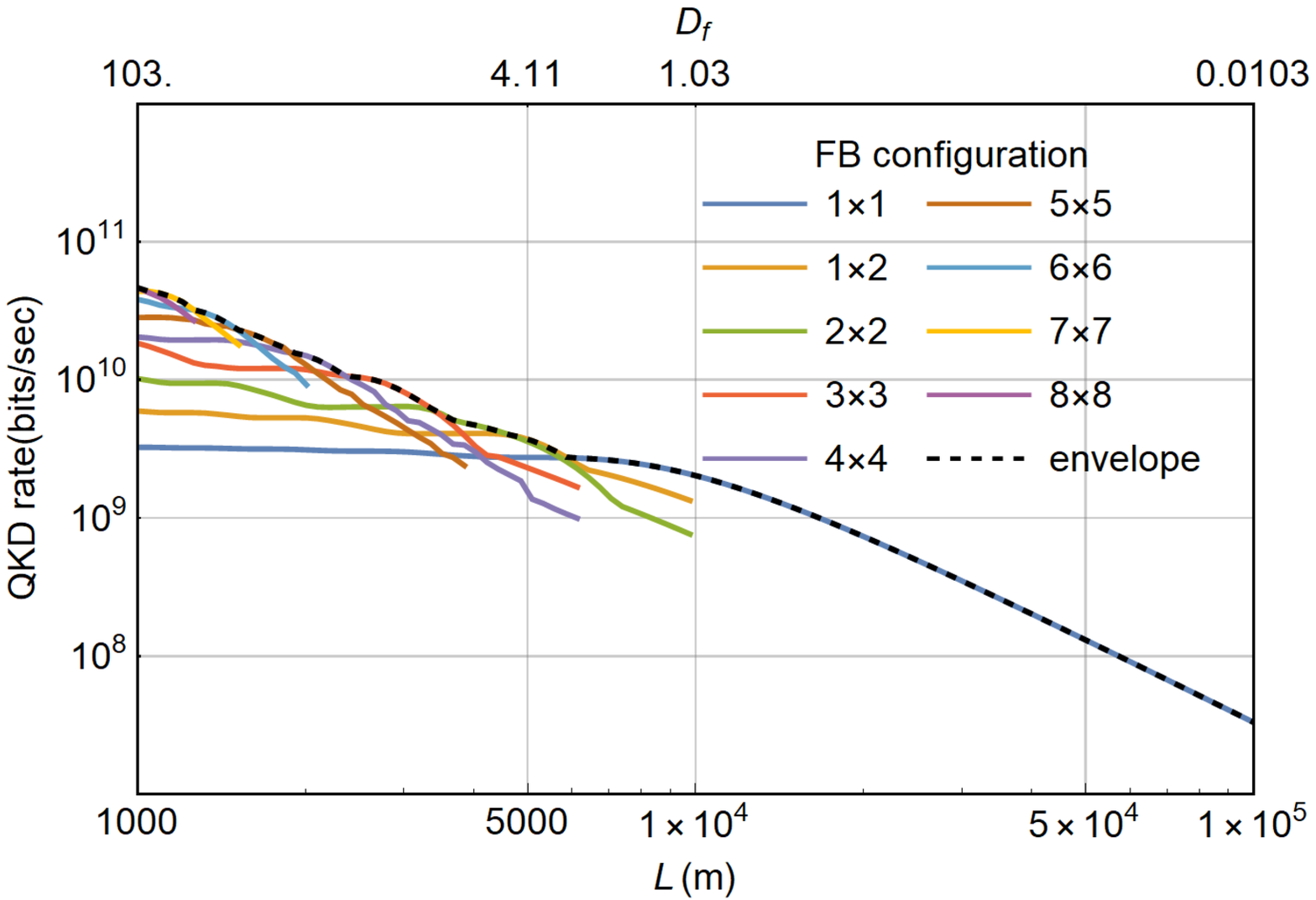}
		\caption{FB modes.}\label{fig:vacuum-FTFB-nonorthogonal}		
	\end{subfigure}
	\hfill
	\begin{subfigure}[t]{0.6\linewidth}
		\centering
		\includegraphics[width=\linewidth]{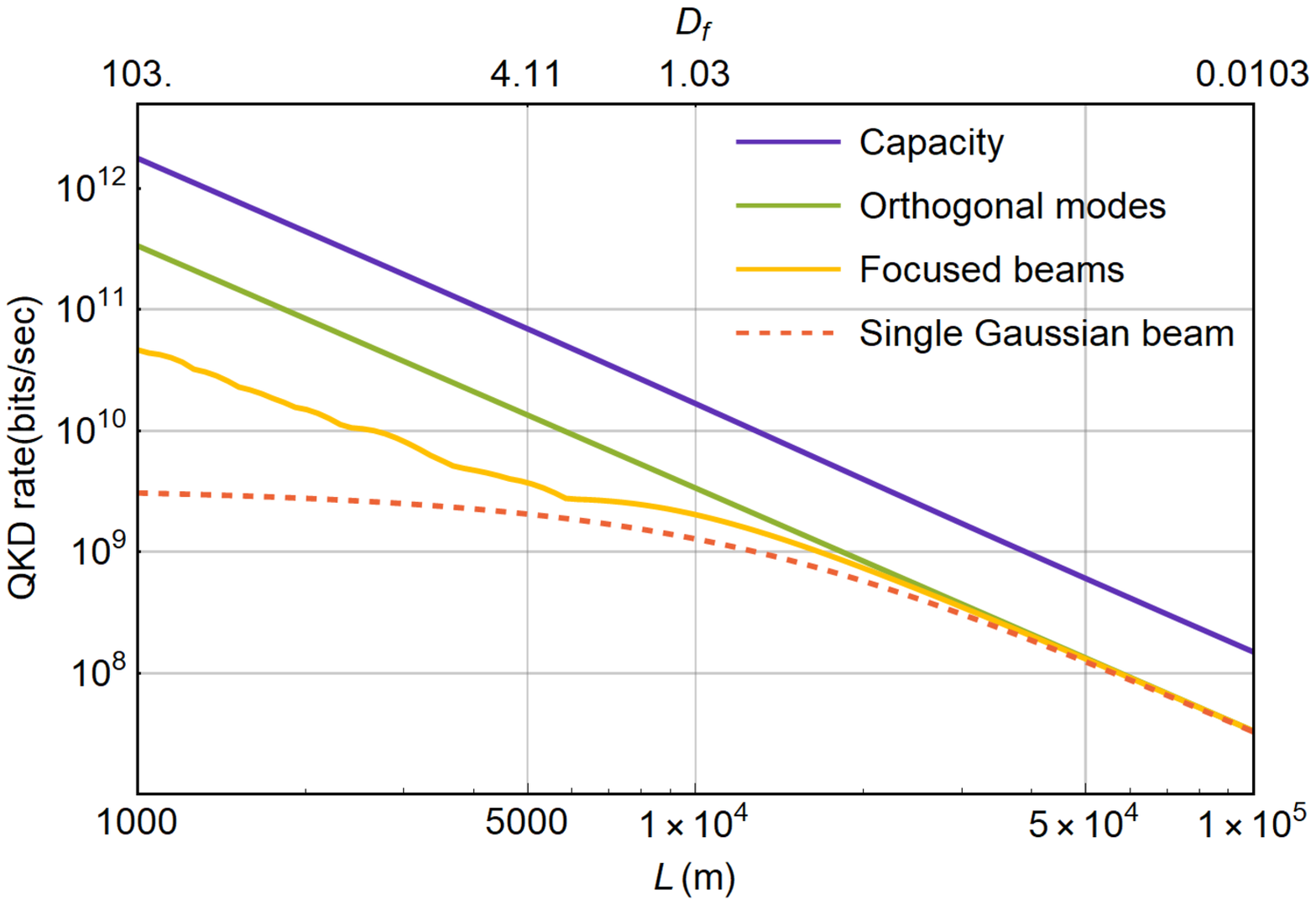}
		\caption{Rate comparison.}\label{fig:vacuum-QKD-rate}		
	\end{subfigure}
		\caption{QKD rates  vs.~path length $L$ for vacuum-propagation channels. In (\subref{fig:vacuum-LG-orthogonal}), the contribution of the high-order LG modes is negligible due to their low transmissivity.  The optimal configuration for the FB modes in (\subref{fig:vacuum-FTFB-nonorthogonal}) depends on the path length due to cross-talk.  Computational limitations make irregular the shapes of the curves in (\subref{fig:vacuum-FTFB-nonorthogonal}).  The comparison of the QKD rates in (\subref{fig:vacuum-QKD-rate}) includes the envelopes for the LG and FB mode sets, as well as the rate achievable using only the focused Gaussian beam and the QKD capacity bound.
				\vspace{-10pt}}\label{fig:vacuum-full-QKDrate}
				\end{figure}

The QKD rate for the orthogonal LG modes described in Section \ref{sec:gaussian_pupils} is 
\begin{align}
\mathcal{R}_{\rm vac,LG}&=\nu\sum_{\bm q} \max_{P_{\bm q}^{\rm (T)}}\mathcal{R}_{\rm QKD}\left(\eta_{\bm q}^{\rm (vac)},P_{\bm q}^{\rm (T)},0\right),
\end{align}
where $\eta_{\bm q}^{\rm (vac)}$ is given in \eqref{eq:gen_modes}.
Although there are infinite number of orthogonal modes, their transmissivities decrease exponentially with $q$, resulting in a negligible contribution to $\mathcal{R}_{\rm vac,G}$ from the high-order modes.
Figure \ref{fig:vacuum-full-QKDrate}(\subref{fig:vacuum-LG-orthogonal}) illustrates this behavior.

The QKD rate for the flat-top focused beams described in Section \ref{sec:square_pupils} is 
\begin{align}
\label{eq:RvacFB}
\mathcal{R}_{\rm vac,FB}&=\nu\max_N \max_{\bm{P}^{\rm (T)}}\sum_{n=1}^{N}\sum_{m=1}^{N} \mathcal{R}_{\rm QKD}\left(\eta_{(n,m)\to(n,m)}^{\rm (vac,FB)},P_{(n,m)}^{\rm (T)},P_{(n,m)}^{\rm (C)}\right),
\end{align} 
where $\bm{P}^{\rm (T)}$ is the vector of power levels transmitted on each focused beam, $P_{(n,m)}^{\rm (C)}$ is the cross-talk power 
\begin{align}
\label{eq:PCvacFB}
P_{(n,m)}^{\rm (C)}&=\sum_{\myatop{n^\prime=1}{n^\prime\neq n}}^{N}\sum_{\myatop{m^\prime=1}{m^\prime\neq m}}^{N}P_{(n^\prime,m^\prime)}^{\rm (T)}\eta_{(n^\prime,m^\prime)\to(n,m)}^{\rm (vac,FB)}
\end{align}
and $\eta_{(n,m)\to(n^\prime,m^\prime)}^{\rm (vac,FB)}$ is given in \eqref{eq:etaFBvac}.
Cross-talk necessitates the optimization of total QKD rate over the power level vector $\bm{P}^{\rm (T)}$.
Furthermore, it limits the number of useful focused beams.
Figure \ref{fig:vacuum-full-QKDrate}(\subref{fig:vacuum-FTFB-nonorthogonal}) shows that the optimal $N$ varies with path length, with 64 focused beams supported in the near-field, and only one in the far-field.


In Figure \ref{fig:vacuum-full-QKDrate}(\subref{fig:vacuum-QKD-rate}) we compare the envelope of the QKD rates  achievable using various focused beam configurations $\mathcal{R}_{\rm vac,FB}$ from Figure \ref{fig:vacuum-full-QKDrate}(\subref{fig:vacuum-FTFB-nonorthogonal}) with the achievable QKD rate $\mathcal{R}_{\rm vac,G}$ using the orthogonal LG modes.
We also report the QKD rate achievable using only the focused Gaussian beam $\Phi_{{\bm 0}}^{\rm (LG)}$ with soft Gaussian pupils and the QKD capacity bound for a single-mode lossy bosonic channel \cite{pirandola17QKDcap} applied to each of the LG modes:
\begin{align}
\mathcal{C}_{\rm QKD}&=-\nu\sum_{\bm q} \log_{2}\left(1-\eta_{\bm q}^{\rm (vac)}\right).
\end{align}


Figure \ref{fig:vacuum-full-QKDrate}(\subref{fig:vacuum-QKD-rate}) illustrates that, despite their limitations, flat-top focused beams capture a significant fraction of the multiplexing gain that is theoretically-achievable by the orthogonal spatial modes.

\subsection{QKD over turbulent channels}
\label{sec:turbulentQKD}
We employ the Kolmogorov-spectrum turbulence model described in Section \ref{sec:prerequisites}.
We obtain the approximate QKD rate for focused beams $\mathcal{R}_{\rm FB}(C_n^2)$, which is now a function of turbulence strength parameter $C_n^2$, by substituting the transmissivity  $\langle\eta_{(n,m)\to(p,l)}^{\rm (FB)}\rangle$ averaged over turbulence from \eqref{eq:etaturbFB} for $\eta_{(n,m)\to(n^\prime,m^\prime)}^{\rm (vac,FB)}$ into \eqref{eq:RvacFB} and \eqref{eq:PCvacFB}.
The approximate QKD rate for the LG modes in turbulence is:
\begin{align}
\label{eq:RturbLG}
\mathcal{R}_{\rm LG}(C_n^2)&=\nu \max\left[\max_Q \max_{\bm{P}^{\rm (T)}}\sum_{{\bm q}:q\leq Q}\mathcal{R}_{\rm QKD}\left(\langle\eta_{\bm{q}\bm{q}}^{\rm (LG)}\rangle,P_{\bm{q}}^{\rm (T)},\langle P_{\bm{q}}^{\rm (C)}\rangle\right),\right.\nonumber\\ 
&\phantom{=\nu \max\left[\vphantom{{{\bm q}:q\leq Q}}\right.}\left.\vphantom{\sum_{{\bm q}:q\leq Q}}\max_{P_0^{\rm (T)}}\mathcal{R}_{\rm QKD}\left(\langle\eta_{{\bm 0}\to\mathrm{R}}^{\rm (LG)}\rangle,P_0^{\rm (T)},0\right)\right],
\end{align}
where $q=2p+|l|+1$ is the order for LG mode $\bm{q}\equiv(p,l)$, $\bm{P}^{\rm (T)}$ is the vector of power transmitted on each mode, $\langle P_{\bm{q}}^{\rm (C)}\rangle$ is the average cross-talk power 
\begin{align}
\label{eq:PCvacLG}
\langle P_{\bm{q}}^{\rm (C)}\rangle&=\sum_{\myatop{{\bm q}^\prime:q^\prime \leq Q}{{\bm q}^\prime\neq {\bm q}}}P_{\bm{q}^\prime}^{\rm (T)}\langle\eta_{\bm{q}^\prime\bm{q}}^{\rm (LG)}\rangle,
\end{align}
$\bm{q}^\prime\equiv(p^\prime,l^\prime)$, $q^\prime=2p^\prime+|l^\prime|+1$, and the calculation of $\langle\eta_{\bm{q}\bm{q}^\prime}^{\rm (LG)}\rangle$ is discussed in Section \ref{sec:gaussian_pupils}.
The outer maximum in \eqref{eq:RturbLG} ensures that power-in-bucket expression \eqref{eq:etaG_PIB} is used in the far-field instead of mode-sorting.
As in the focused beam setup, the turbulence-induced cross-talk limits the number of useful LG modes, and the optimal mode order $Q$ varies with path length.
Furthermore, total QKD rate must be optimized over the power vector $\bm{P}^{\rm (T)}$.

\begin{figure} 
    \centering
\begin{subfigure}[t]{0.49\linewidth}
		\centering
		\includegraphics[width=\linewidth]{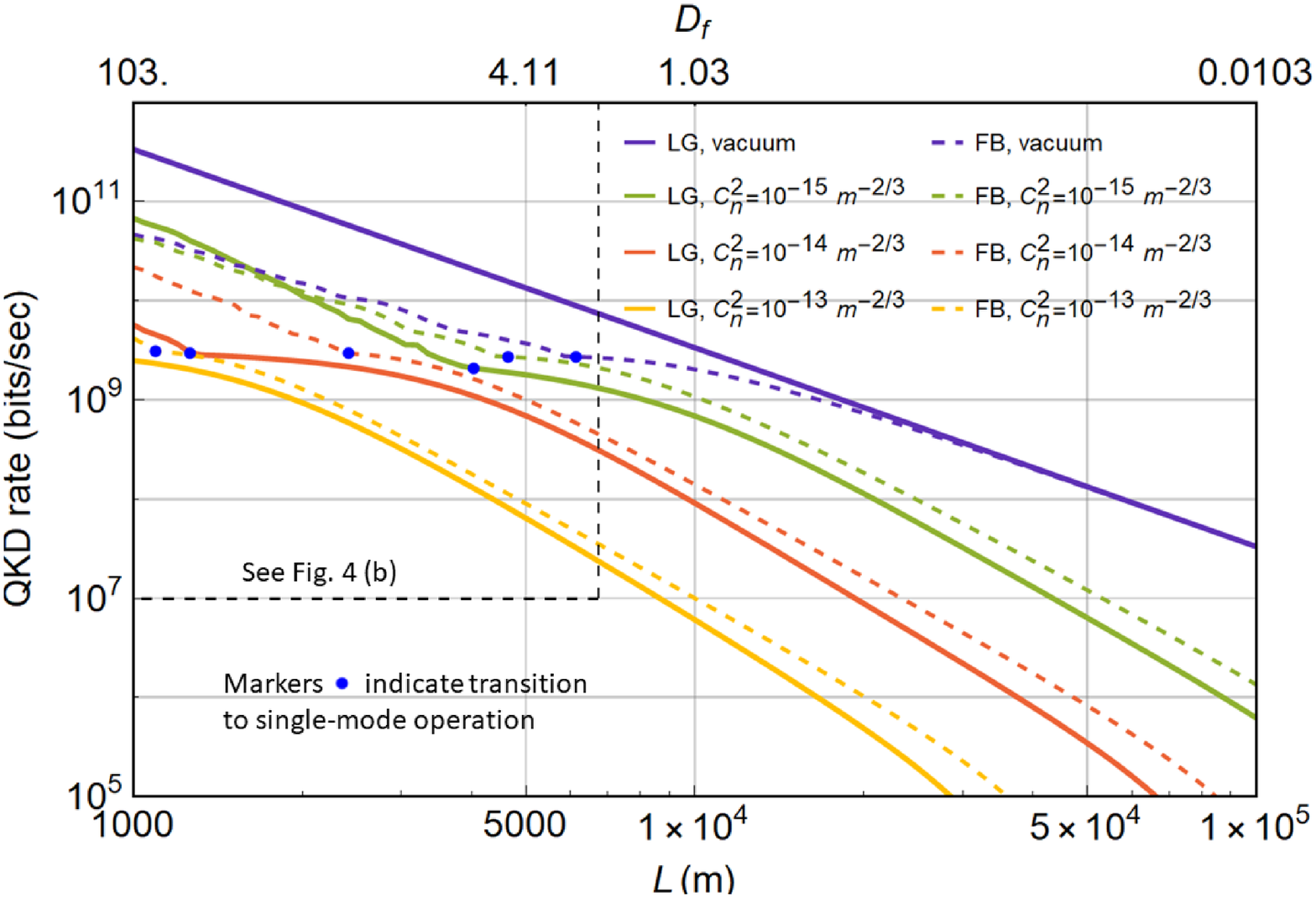}
		\caption{QKD rates over $L\in[1,100]$ km path length. The irregular shapes of the curves is due to their being rate envelopes for different systems, and computational limitations.}\label{fig:QKDrate_all}		
	\end{subfigure}
	\hfill
\begin{subfigure}[t]{0.49\linewidth}
		\centering
		\includegraphics[width=\linewidth]{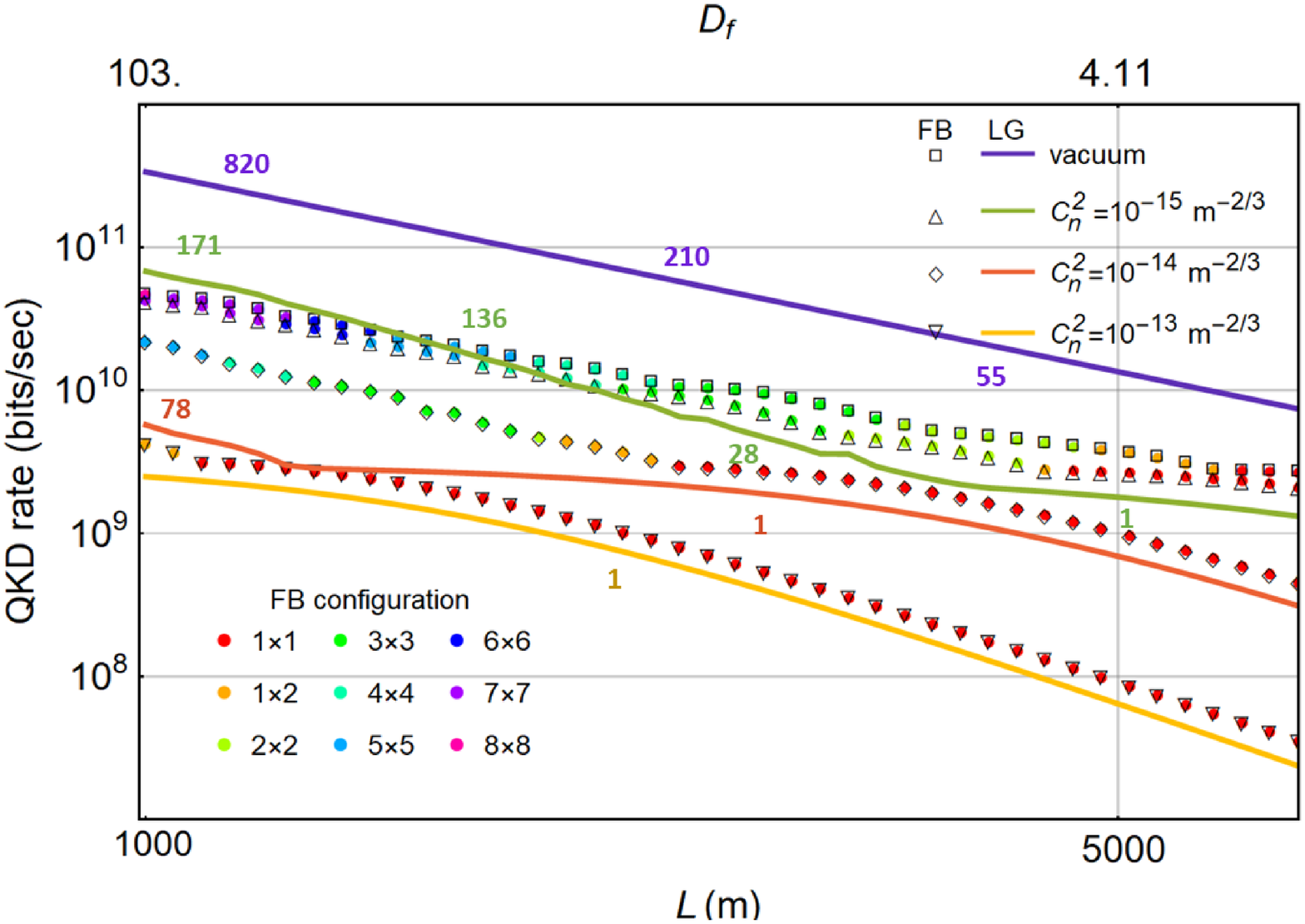}
		\caption{Near-field QKD rates. Marker colors indicate optimal FB configurations, numbers near LG curves indicate number of modes used (optimized for propagation in turbulence).}\label{fig:QKDrate_nearfield}		
	\end{subfigure}
		\caption{Comparison of the vacuum- and turbulent-propagation QKD rates for the LG and FB mode sets vs.~path length $L$. 
				\vspace{-10pt}}\label{fig:QKDrate}
				\end{figure}

We compare these approximate QKD rates in turbulence with LG modes and focused beams in Figure \ref{fig:QKDrate}.
The rates are evaluated in the weak ($C_n^2=10^{-15}~\text{m}^{-2/3}$ ), medium ($C_n^2=10^{-14}~\text{m}^{-2/3}$), and strong ($C_n^2=10^{-13}~\text{m}^{-2/3}$) turbulence regimes, and are also compared to the vacuum scenario.
The change in rate scaling from $L^{-2}$ in vacuum to $L^{-16/5}$ in turbulence is clearly evident in the far field.
Surprisingly, our flat-top focused beam architecture outperforms the LG mode set in all but near-field weakly-turbulent scenarios, even though we assumed availability of perfect mode generation and sorting for the latter.
Furthermore, achieving maximum rate seems to need substantially more LG modes than FBs.
Thus, even if mode sorting were free, the optimal operation of LG-mode system requires requires a significantly greater number of costly single-photon detectors and associated electronics than that for our proposed FB system.

\section{Conclusion}
Generation and separation of the LG modes is an active research area in optics. However, despite the significant technological advances that reduce the device size, weight, and cost, our results show their apparent inefficiency compared to potentially simpler focused beams in practical QKD applications.
Thus, further investigation of QKD using multiple overlapping focused beams is necessary, including the optimization of the pixel layout, as well as the emulation and experiments involving physical platforms.

\appendix

\section*{Appendix: Comparison of the $5/3$-law and the square-law approximation in power transfer calculations}\label{wave-structure-function}
\begin{figure}[htbp]
\centering
\includegraphics[width=0.6\linewidth]{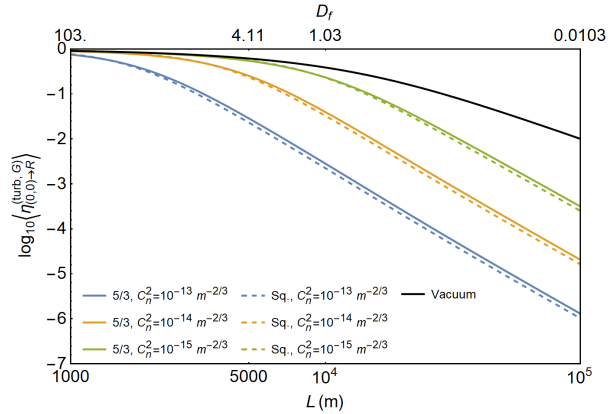}
\caption{Comparison between the $5/3$-law turbulence structure function and its square-law approximation in the evaluation of the average power-in-bucket transmissivity for the Gaussian beam $\eta_{{\bm 0}\to\mathrm{R}}^{\rm (LG)}$.  Similar results were obtained for other geometries.} 
\label{fig:squareX53}
\end{figure}

The square-law approximation \eqref{eq:D} to the $5/3$-law Kolmogorov-spectrum turbulence structure function offers substantial reduction in complexity of the calculations performed here, and in other works \cite{chandrasekaran14pieturb1,chandrasekaran14pieturb2,ghost-imaging}.
Figure \ref{fig:squareX53} compares the average power-in-bucket transmissivity for the Gaussian beam $\eta_{{\bm 0}\to\mathrm{R}}^{\rm (LG)}$ evaluated using the $5/3$-law and square-law approximation. 
We employed soft Gaussian pupils with $R=10$ cm at operating wavelength $\lambda=1.55 \mu$m.
Figure \ref{fig:squareX53} demonstrates that, while the square-law approximation underestimates the power transmissivity, it is fairly accurate.
We obtained similar results for other channel geometries.

\section*{Acknowledgments}
WH and SG acknowledge Office of Naval Research (ONR) contract number N00014-19-1-2189, and the NSF Center for Quantum Networks (CQN) awarded under grant number 1941583. 
\section*{Disclosures}
The authors declare no conflicts of interest.

\bibliography{QKD_papers}

\end{document}